\documentclass{article}

\usepackage{xcolor}
\usepackage{spconf}
\usepackage{graphicx}
\usepackage{amsmath}
\usepackage{bm}
\usepackage{amssymb}
\usepackage{booktabs}
\usepackage{multirow}
\usepackage{diagbox}
\usepackage{array}
\usepackage[hang,flushmargin]{footmisc}
\usepackage[hyphens]{url}

\title{CasNet: Investigating Channel Robustness for Speech Separation}

\name{Fan-Lin Wang$^2$, Yao-Fei Cheng$^2$, Hung-Shin Lee$^1$, Yu Tsao$^3$, and Hsin-Min Wang$^2$}
\address {
$^1$North Co., Ltd., Taiwan \\
$^2$Institute of Information Science, Academia Sinica \\
$^3$Research Center for Information Technology Innovation, Academia Sinica
}
\begin{document}
%
\maketitle
\begin{abstract}
Recording channel mismatch between training and testing conditions has been shown to be a serious problem for speech separation. This situation greatly reduces the separation performance, and cannot meet the requirement of daily use. In this study, inheriting the use of our previously constructed TAT-2mix corpus, we address the channel mismatch problem by proposing a channel-aware audio separation network (CasNet), a deep learning framework for end-to-end time-domain speech separation. CasNet is implemented on top of TasNet. Channel embedding (characterizing channel information in a mixture of multiple utterances) generated by Channel Encoder is introduced into the separation module by the FiLM technique. Through two training strategies, we explore two roles that channel embedding may play: 1) a real-life noise disturbance, making the model more robust, or 2) a guide, instructing the separation model to retain the desired channel information. Experimental results on TAT-2mix show that CasNet trained with both training strategies outperforms the TasNet baseline, which does not use channel embeddings.
\end{abstract}
\section{Introduction}
\label{sec:intro}

Speech separation \cite{Wang2018} originates from the cocktail party problem \cite{Haykin2005}, which refers to the perception of each speech source in a noisy social environment. To understand each speaker's speech, we first need to separate overlapping speech, which is the goal of speech separation. As a necessary pre-processing for downstream tasks, such as speaker diarization \cite{Neumann2019} and automatic speech recognition \cite{Raj2020a}, many efforts have been made in speech separation.

Nowadays, the main dataset used in speech separation research is the WSJ0-2mix dataset \cite{Hershey2016}. In WSJ0-2mix, an artificially synthesized dataset, all mixed utterances are full overlaps of clean speech from two speakers. In recent research, a popular architecture is the time-domain audio separation network (TasNet) \cite{Luo2018}. Many TasNet-based models have achieved extraordinary performance \cite{Luo2020,Chen2020,Zeghidour2020,Hu2021,Subakan2021} on WSJ0-2mix. However, WSJ0-2mix sets many restrictions on the experiments, which may lead to domain mismatches. 

Domain mismatch can be attributed to four factors: speaker, content, channel, and environment. Regarding speaker mismatch, the speakers in the test sets of all datasets are designed to be unseen in the training sets. However, there is no noticeable drop in performance, demonstrating the speaker generalization of these models. Environment mismatch refers to reverberation and noise that may be encountered in reality and are not seen in the training set. To address this issue, two new datasets have been presented: WHAM! \cite{Wichern2019} and WHAMR! \cite{Maclejewski2020}, which are the noisy and reverberant extensions of WSJ0-2mix, respectively.

Content mismatch focuses on what the speaker said, such as vocabulary or even different languages that contain various phonemes. In \cite{Kadioglu2020,cosentino2020}, the authors argue that the larger the vocabulary presented in the training set, the better the generalization of the model. In \cite{Borsdorf2021_cock}, using the GlobalPhoneMS2 dataset consisting of 22 spoken languages, the authors show that when trained on a multilingual dataset, the model can improve its performance on unseen languages. Regarding channel mismatch, it focuses on the type of microphone used in the recording. The authors of \cite{Maciejewski2019} argue that near-field data are easier to separate than far-field data, even though both were recorded in the same environment. 

In the COVID-19 pandemic era, virtual meetings have become prevalent and recorded with a wider variety of microphones. Furthermore, smartphones are frequently used tools in the daily recording. If all training sets are recorded with condenser microphones, the speech separation performance will drop significantly in daily use. Therefore, channel mismatch should be investigated in more depth to meet demand. In our previous work \cite{Wang2022}, we found that the impacts of different languages are small enough to be ignored compared to the impacts of different channels. Also, although the content is the same, the separation performance varies due to the different microphones. To address the channel mismatch, it is necessary to create a channel-robust speech separation model. Here, we define ``channel robustness'' in two directions according to the channel of the target. The first is classic speech separation: no matter which channel the mixture is recorded through, the separated utterances must remain on the same channel as the mixture. The other definition is that separated utterances should be enhanced as if they were recorded by a clean channel, i.e., a condenser microphone. The benefit of this definition is that the downstream model does not need to be channel-robust when receiving the output of the separation model. In both definitions, the separation model should perform well on channels unseen in the training set.

In this paper, we focus on the first definition of channel robustness. We propose a channel-aware audio separation network, CasNet, which can separate mixtures guided by channel embeddings. A channel encoder inspired by speaker verification models is designed to generate channel embeddings, which are introduced into the separation module via the FiLM technique. Our model can be applied to any TasNet-based model to enhance channel robustness. We conducted experiments on the TAT-2mix dataset designed in our previous work and explored the role of channel embeddings as guiding or disturbing by inputting different auxiliary mixtures during model training. The results show that CasNet trained in both ways outperform the TasNet baseline. We open-sourced the code for training on GitHub\footnote{\url{https://github.com/Sinica-SLAM/CasNet}}.

The contribution of this paper spans the following aspects: 1) To the best of our knowledge, we are the first to study solutions for channel mismatch; 2) we create a module that can generate channel embeddings to enhance robustness; 3) we investigate the different roles of channel embeddings. 4) our proposed model, CasNet, outperforms the TasNet baseline.

\section{Proposed: CasNet}
\label{sec:model}

\begin{figure}[t]
\centering
\includegraphics[width=0.49\textwidth]{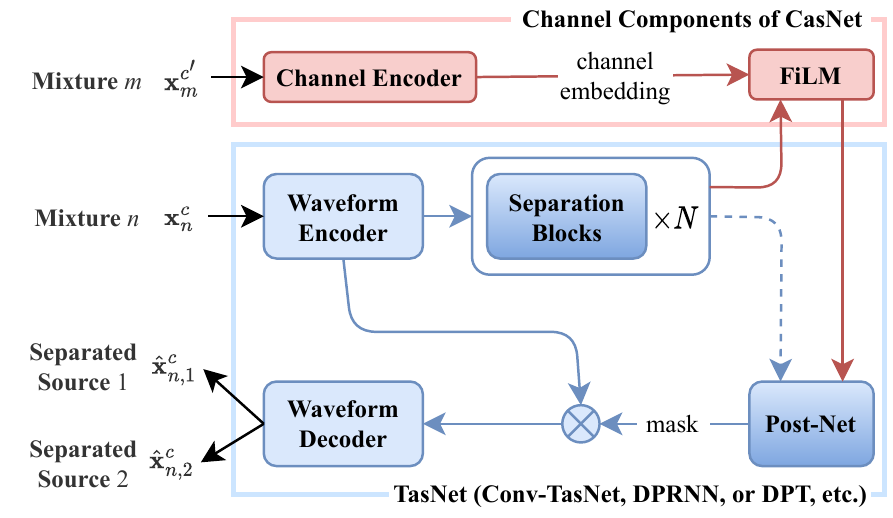}
\vspace{-15pt}
\caption{The architecture of CasNet. The blue blocks are the original TasNet modules. The red blocks are our new design.}
\label{fig:casnet}
\vspace{-15pt}
\end{figure}

As shown in Fig. \ref{fig:casnet}, our model is based on TasNet but with a channel encoder. The input of TasNet is a mixture of overlapping speech $\mathbf{x}_{n}^{c}$, and the outputs of TasNet are separated utterances $\mathbf{\hat{x}}_{n,i}^{c}$. Using an auxiliary mixture $\mathbf{x}_{m}^{c'}$, the channel encoder generates a channel embedding to support or interfere with separation. The separation model integrates the channel embedding after the separation blocks and before the Post-Net. During training, we considered different schemes whether $m = n$ or not and whether $c = c'$ or not. In the inference stage, we examined different $\mathbf{x}_{m}^{c'}$ to feed in the Channel Encoder, as shown in Fig. \ref{fig:test_stage}. The details of each module are described as follows. Training objectives will be discussed at the end of this section.

\subsection{TasNet}
\label{ssec:tasnet}

The main concept of TasNet is speech separation in the time domain. The waveform input is not transformed into a spectrogram for processing, and the output is also a waveform. TasNet is mainly composed of three parts: waveform encoder, separator, and waveform decoder. First, the waveform encoder (Conv1D) takes the mixture as input and transforms it into the corresponding representation. This representation is then fed into the separator, which estimates individual speaker masks. Finally, the waveform decoder (TransposeConv1D) is used to reconstruct each source waveform from the masked encoder features.

The main development of TasNet has focused on the structure of the separator. The separator consists of a stack of separation blocks and a Post-Net. The separation block evolved from a fully-convolutional module in Conv-TasNet \cite{Luo2019} to a Transformer-based module in Sepformer \cite{Subakan2021}. Along with the improvement in performance, the Post-Net has also become more complex. All details can be found in the individual original papers.

\subsection{Channel Encoder}
\label{ssec:channelnet}

Inspired by previous speaker verification research \cite{Desplanques2020,Chung2020a,Rouvier2021}, the structure of Channel Encoder is mainly composed of a Residual Net and a pooling layer. We argue that the input should be a mixture instead of a single-speaker utterance to ensure that the channel embedding mainly captures channel information rather than speaker information. Suppose a training batch $\{\mathbf{x}_{m}^{c'}\}^{M}_{m=1}$ contains $M$ input utterances. First, the batch is transformed into a representation $\mathbf{X}_0$ in the embedding space by the same waveform encoder as TasNet. Then, $\mathbf{X}_0$ is sent to the Residual Net. 

The Residual Net is composed of one ConvBlock and $B$ SE-ResBlocks. $\mathbf{X}_0$ is processed by the ConvBlock, which consists of a 1D convolution ($Conv1D$), a nonlinear activation function ($ReLU$), and a normalization operation ($BatchNorm$) \cite{Ioffe2015}, as expressed by
\begin{equation} \label{eq_resblock}
 \mathbf{X}_{1} = BatchNorm(ReLU(Conv1D(\mathbf{X}_{0}))).
\end{equation}
For $i=1,\ldots,B$, $\mathbf{X}_{i}$ is fed into SE-ResBlocks. In a SE-ResBlock, there are two ConvBlocks and a squeeze-and-excitation (SE) layer. The SE layer utilizes a two-layer fully connected network ($FC$) with average pooling ($AvgPool$) and the sigmoid function ($Sigmoid$) to calculate the weights of the original feature maps and scales each dimension of the channel according to its importance. The SE process involves two steps: 1) generating global information (squeeze step); and 2) re-weighting each feature map (excitation step), as recursively expressed by
\begin{equation} \label{eq_se}
\mathbf{X}_{i+1} = Sigmoid(FC(AvgPool(\mathbf{X}_{i}))) \times \mathbf{X}_{i}+\mathbf{X}_{i},
\end{equation}
where $i$ ranges from 1 to $B-1$. Note that after each SE-ResBlock, a residual path is added to the end of the block.

Then, the output $\mathbf{X}_{B}$ goes through a pooling layer to get the channel embedding.
An attentive pooling layer is used to compute the weighted mean of the last dimension (i.e., time frames) of $\mathbf{X}_{B}$: 
\begin{equation} \label{eq_alpha}
\mathbf{A}=Sigmoid(FC(\mathbf{X}_{B})),\:\mathbf{A}=[\mathbf{a}_1,\dots,\mathbf{a}_{M}],
\end{equation}
\begin{equation} \label{eq_pool}
\mathbf{Z} = \mathbf{A}^T \times \mathbf{X}_B.
\end{equation}
where $\mathbf{A}\in\mathbb{R}^{T \times M}$, and $T$ is the temporal length. Finally, a linear operation ($Linear$) is applied to produce the channel embedding $\mathbf{C}=[\mathbf{c}_1,\dots,\mathbf{c}_{M}] \in \mathbb{R}^{D \times M}$, where $D$ is the dimension of channel embedding:
\begin{equation} \label{eq_pool_relu}
\mathbf{C} = Linear(\mathbf{Z}).
\end{equation}

\subsection{FiLM}
\label{ssec:film}

We adopt FiLM \cite{Perez2018} to integrate channel embedding into TasNet. We first transform the channel embedding $\mathbf{C}$ to weight $\mathbf{W}$ and bias $\mathbf{b}$ by two different linear operations:
\begin{equation} \label{cc}
 \mathbf{W} = Linear(\mathbf{C}),\:\mathbf{b} = Linear(\mathbf{C}).
\end{equation}
The separation feature maps $\mathbf{S}$ estimated by TasNet (waveform encoder + $N$ separation blocks) is normalized by instance normalization ($Norm$) \cite{Ulyanov2016} and multiplied by the weight $\mathbf{W}$ and biased by $\mathbf{b}$, and then goes through a nonlinear activation function ($PReLU$): 
\begin{equation} \label{prelu}
 \mathbf{S}' = PReLU(\mathbf{W} \times Norm(\mathbf{S}) + \mathbf{b}).
\end{equation}
Finally, the separation feature maps $\mathbf{S}'$ are turned into the estimated masks by the Post-Net.

\subsection{Training Objectives}
\label{ssec:train}

The main training objective for speech separation is to minimize the reconstruction loss $\mathcal{L}_{rc}$ according to the scale-invariant signal-to-distortion ratio (SI-SNR). SI-SNR can be calculated with the output of the network and the target source through the definition in Eqs. 13-15 of \cite{Luo2018}.

Our goal is to maximize SI-SNR or minimize the negative SI-SNR. Since there are multiple speakers in a mixture, permutation-invariant training (PIT) is performed, which computes the combination that yields the highest SI-SNR for backward propagation. Additionally, to ensure that the channel embedding contains channel information, we include the channel identification loss $\mathcal{L}_{ci}$, which is a cross-entropy loss. We feed the channel embedding into a channel classifier to distinguish through which channel $\mathbf{x}_{m}^{c'}$ is recorded. Therefore, the total loss used in this paper is 
\begin{equation}
 \mathcal{L}_{total}=\mathcal{L}_{rc}+\gamma\times\mathcal{L}_{ci},
\end{equation}
where $\gamma$ is the weight of the channel identification loss.

\begin{figure}[t]
\centering
\includegraphics[width=0.48\textwidth]{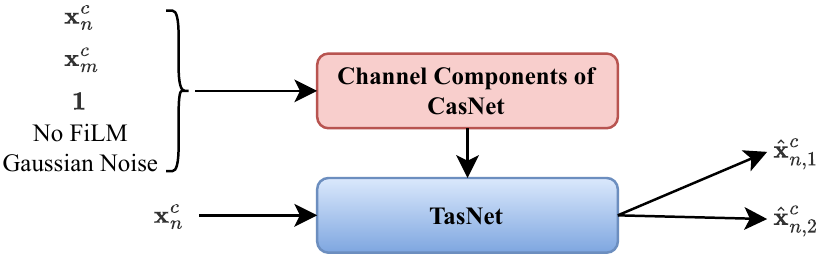}
\vspace{-20pt}
\caption{The inference stage of CasNet.}
\label{fig:test_stage}
\vspace{-10pt}
\end{figure}

\section{Experiments}
\label{sec:experiments}

\subsection{Dataset}
\label{ssec:dataset}

The dataset used in this paper is TAT-2mix, a dataset created in our previous work \cite{Wang2022}. TAT-2mix is based on the Taiwanese across Taiwan (TAT) corpus \cite{Liao2020}. Taiwanese, also known as Southern Min, is a common dialect in Taiwan and belongs to the same language family as Mandarin. The corpus was recorded on six channels simultaneously, including one close-talk (Audio-Technica AT2020), one distant X-Y stereo microphone (ZOOM XYH-6 stereo microphone, containing left and right channels), one lavalier (Superlux WO518+PS418D), iOS devices (including iPhones, iPads, and iPods), and Android phones (produced by ASUS and Samsung). From these six channels, we created six corresponding datasets. All datasets are comprised of the same mixtures of utterances but from different channels. TAT-2mix is designed to have the same number of mixed utterances and similar statistics as WSJ0-2mix. 
Details can be referred to in our previous work \cite{Wang2022}.

\subsection{Experimental Setup}
\label{ssec:configuration}

All our experiments were performed using the SpeechBrain toolkit \cite{Ravanelli2021}. DPRNN-TasNet was used as the TasNet baseline. CasNet was implemented on the TasNet baseline by adding the proposed channel embedding mechanism. We used the Adam optimizer with default parameters. The learning rate starts at 1.5e-4 and is halved when there is no improvement for 2 epochs. There are 4 SE-ResBlocks ($B = 4$) in the Residual Net of CasNet, and the dimension of channel embedding is 128 ($D = 128$). CasNet was trained for 100 epochs on 3-second-long segments with a batch size of 8. 

There are six types of channels in TAT-2mix. We trained with five of them and tested the models with the sixth channel to test the channel robustness. In our previous work \cite{Wang2022}, we found that training with $\textnormal{TAT-2mix}_\textnormal{Android}$ can achieve a certain level of channel robustness, since there are multiple brands in the Android system, including ASUS and Samsung. Therefore, we chose the Android channel as the test channel to avoid gaining robustness from a single kind of channel. In each minibatch, we randomly selected a channel as ``$c$'' for training to save training time.

\subsection{Results}
\label{ssec:results}

According to the selection of $\mathbf{x}_{m}^{c'}$ during model training, our experiments are divided into two parts. In the first experiment, the channel of $\mathbf{x}_{m}^{c'}$ is the same as that of $\mathbf{x}_{n}^{c}$ (i.e., $c' = c$) to investigate the guiding ability of channel embedding. Inspired by target speaker extraction \cite{Zmolikova2019}, where models use a speaker vector to extract a target speaker in a mixture, channel embeddings can be considered a guide in speech separation to extract the signal of the target channel. In the second experiment, training $\mathbf{x}_{m}^{c'}$ and $\mathbf{x}_{n}^{c}$ have different content and are from different channels (i.e., $m \neq n$ and $c \neq c'$). We believe that CasNet can become more robust when different interfering mixtures are used in training, similar to how data augmentation adds interference to the training data to make speaker verification models robust.

\begin{table}[t]
\caption{Speech separation performance on TAT-2mix when training $\mathbf{x}_{m}^{c'}$ and $\mathbf{x}_{n}^{c}$ are from the same channel ($c = c'$). Whether the content of $\mathbf{x}_{m}^{c'}$ is the same as that of $\mathbf{x}_{n}^{c}$ during training ($m = n$) is noted in the \textbf{utter.} column. The number of parameters is recorded in the \textbf{param.} column. \boldsymbol{$\gamma$} is the weight of the channel identification loss. }
\vspace{5pt}
\label{tab:guide}
\small
\centering
\setlength{\tabcolsep}{9pt}
\begin{tabular}{lcccc}
\toprule
\textbf{Model} & \textbf{param.} & \textbf{utter.} & \boldsymbol{$\gamma$} & \textbf{SI-SNR (dB)} \\ 
\midrule\midrule
Baseline & 14.6M & - & - & 9.16 \\
Large Base & 19.4M & - & - & 9.72 \\
\midrule
\multirow{4}{*}{CasNet} & \multirow{4}{*}{17.1M} & \multirow{4}{*}{same} & 1 & 9.25 \\
 & & & 0.1 & 9.63 \\
 & & & 0.01 & 9.64 \\
 & & & 0 & \textbf{9.76} \\
\midrule
CasNet & 17.1M & diff & 0 & 9.09 \\
\midrule
Topline & 14.6M & - & - & \textbf{10.55} \\
\bottomrule
\end{tabular}
\vspace{-10pt}
\end{table}

The results of the first experiment are shown in Table \ref{tab:guide}. The number of parameters is recorded in the \textbf{param.} column. Whether $m = n$ or not during training is noted in the \textbf{utter.} column. In the inference phase, the input of TasNet and Channel Encoder is the same. We trained the TasNet baseline on five channels and the TasNet topline on all six channels. By comparing the baseline and topline, we can see that training on multiple channels can already achieve some degree of channel robustness, even though the test channel is not seen during training, which is consistent with the results in \cite{Borsdorf2021_cock}. Since CasNet has more parameters than the TasNet baseline, we also trained a large TasNet model by adding two more separation blocks for a fair comparison. Although the large TasNet model has more parameters than CasNet, its performance is worse than the best performance of CasNet, indicating that the success of CasNet is not due to the additional model capacity, but the design of the structure. For CasNet, the results for the $m=n$ case seem to suggest that the channel identification loss is not important, as the best performance is achieved with $\gamma=0$. The result of the $m \neq n$ case is not as expected. This requires further investigation, e.g., considering different $\gamma$ values in the $m \neq n$ case.

\begin{table}[t]
\caption{Speech separation performance (SI-SNR, unit: dB) on TAT-2mix when training $\textbf{x}_{m}^{c'}$ and $\textbf{x}_{n}^{c}$ have different content and are from different channels ($c \neq c'$ \& $m \neq n$). The columns \textbf{Train} and \textbf{Test} show the data used for channel embedding extraction. $\textbf{1}$ stands for an all-one vector, and Gaussian means that each element in the channel embedding is a random sample from a standard normal distribution.}
\vspace{5pt}
\label{tab:mislead}
\small
\centering
\setlength{\tabcolsep}{13pt}
\begin{tabular}{cccc}
\toprule
\textbf{Train} & \textbf{Test} & \boldsymbol{$\gamma = 0$} & \boldsymbol{$\gamma = 0.01$} \\ 
\midrule\midrule
\multirow{4}{*}{$\textbf{x}_{m}^{c^\prime}$}
& $\textbf{x}_{n}^{c}$ & 9.22 & \textbf{9.63} \\
& $\textbf{x}_{m}^{c}$ & 9.20 & \textbf{9.63} \\
& $\textbf{1}$ & 7.49 & 7.23 \\ 
& No FiLM & 4.27 & 2.03 \\
& Gaussian Noise & 7.19 & 7.71 \\
\bottomrule
\end{tabular}
\vspace{-10pt}
\end{table}

The results of the second experiment are shown in Table \ref{tab:mislead}. 
In this experiment, during training, $\textbf{x}_{m}^{c'}$ and $\textbf{x}_{n}^{c}$ have different content and are from different channels. In the inference phase, five different channel sources are considered, as illustrated in Fig. \ref{fig:test_stage}. We can see that the CasNet trained with channel embedding perturbation still outperforms the TasNet baseline. We also see that the channel identification loss is crucial in this case, despite its small weight. From the experience of data augmentation, disturbance should be removed in the inference stage. Therefore, we replaced $\textbf{x}_{m}^{c'}$ with an all-one vector or removing FiLM during inference. However, their performance is even worse than that of the TasNet baseline probably due to mismatch between training and testing. Finally, we used a vector whose elements are all random samples from a standard normal distribution as $\textbf{x}_{m}^{c'}$ to prove that the channel embedding should not be a random disturbance, and the results are as expected.

\section{Conclusion and Future Work}
\label{sec:conclusion}

To alleviate the channel mismatch problem, we have proposed a channel-aware audio separation network CasNet, which is based on TasNet with an additional Channel Encoder that generates channel embeddings. The embedding is integrated with TasNet via the FiLM technique. In the experiments, we explored different roles of channel embeddings and found that both training methods, either guiding or disturbing, outperformed the TasNet baseline. In the future, we will investigate the second definition of channel robustness, i.e., enhancing speech with reduced channel effects during separation. 

\vfill\pagebreak

\bibliographystyle{IEEE}
\small\bibliography{references.bib}

\end{document}